# Sustainable Next-generation Color Converters of *P. harmala* Seed Extracts for Solid-State Lighting

*Talha Erdem[1,*], Ali Orenc[2], Dilber Akcan[3], Fatih Duman[2,4], Zeliha Soran-Erdem[5,*]*

[1] Department of Electrical-Electronics Engineering, Abdullah Gül University, Kayseri, Türkiye

[2] Nanotechnology Research Center (ERNAM), Erciyes University, Kayseri, Türkiye

[3] Bioengineering Graduate Program, Abdullah Gül University, Kayseri, Türkiye

[4] Department of Biology, Erciyes University, Kayseri, Türkiye

[5] Department of Engineering Sciences, Abdullah Gül University, Kayseri, Türkiye

**Abstract**

Traditional solid-state lighting relies on color converters with a serious environmental footprint. As an alternative, natural materials such as plant extracts could be employed if their low quantum yield (QYs) in liquid and solid states were higher. With this motivation, here, we investigate the optical features of *P. harmala* extract in water, develop its efficient color-converting solids using a facile, sustainable, and low-cost method, and integrate it with a light-emitting diode. To obtain a high-efficiency solid host for the *P. harmala*-based fluorophores, we optically and structurally compared two crystalline and two cellulose-based platforms. Structural characterizations indicate that sucrose crystals, cellulose-based cotton, and paper platforms allow fluorophores to be distributed relatively homogenously as opposed to the KCl crystals. Optical characterizations reveal that the extracted solution and the extract-embedded paper possess QYs of 75.6% and

44.7%, respectively, whereas the QYs of the cotton, sucrose, and KCl crystals remain below 10%. Subsequently, as a proof-of-concept demonstration, we integrate the as-prepared efficient solid of *P. harmala* for the first time with a light-emitting diode (LED) chip to produce a color-converting LED. The resulting blue-emitting LED achieves a luminous efficiency of 21.9 lm/W$_{elect}$ with CIE color coordinates of (0.139,0.070). With these results, we bring plant-based fluorescent biomolecules to the stage of solid-state lighting. We believe that they hold great promise as next-generation, environmentally friendly organic color converters for lighting applications.

**Keywords:** Plant extract, solid-state lighting, *P. harmala*, color converter, light-emitting diode

**TOC Figure:**

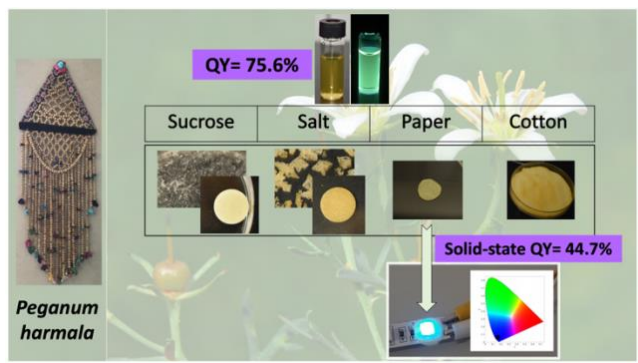

# 1. Introduction

Solid-state lighting (SSL) based on light-emitting diodes (LEDs) offers high-quality lighting with higher efficiency and lower energy consumption in addition to a substantially longer lifetime[1]. Current SSL technology mainly relies on employing a blue-emitting LED chip integrated with color-converting materials[2]. The most widely used material system forming the color converting films is the phosphors made of rare-earth metal ions[3]. Despite their high photoluminescence quantum yields (QY), alternatives are in demand due to 1) the difficulty in tuning their colors limiting their performance, 2) the concerns regarding their supply, and 3) the environmental footprint of mining the rare-earth metals[4]. Among the alternatives, quantum dots (QDs) have stepped forward as they enable easy color-tuning of the LEDs[5] and are less vulnerable to the supply crises[6]. Nevertheless, the efficient QDs are toxic due to their cadmium content while their synthesis is far from being environmentally friendly[7]. Although organic semiconductors could be promising despite their stability issues[8], their syntheses possess a substantial environmental footprint similar to the QDs. In this work, we explore the possibility of using plant extracts as sustainable alternatives to phosphors, QDs, and organic semiconductors, with an emphasis on *Peganum harmala*.

Plants have been traditionally used for curation owing to various biomolecules that they have, such as alkaloids, flavonoids, phenols, anthocyanins, etc.[9–12]. Extraction of these molecules has been studied intensively in the literature, mainly for pharmacologic and therapeutic purposes[12]. On the other hand, some of these plants possess autofluorescent biomolecules which absorb the light and reemit at longer wavelengths[13]. This feature enables imaging certain plant parts without tagging with a reporter molecule[13,14]. However, the potential of these autofluorescent biomolecules has not been evaluated as alternatives to synthetic fluorescent probes including semiconducting polymers, their nanoparticles, or quantum dots (QDs). Despite their availability, sustainability, eco-friendly nature, bio-compatibility, and low cost,[15] a limited number of studies report their use in applications other than pharmacology and biotechnology. These include flexible electronics (e-skins)[16], solar cells[17], and energy storage[18]. For lighting applications, Roy et al.[19] synthesized green-emitting graphene quantum dots from plant extracts obtained from Neem (*Azadirachta*

*indica*) and Fenugreek (*Trigonella foenum-graecum*), and employed them to obtain white LEDs. In another study conducted by Singh and Mishra[20], fluorescent polymer films were prepared using red pomegranate and turmeric extracts to generate white light on a blue LED. In this work, the QYs of the fluorescent molecules in the extract were reported as <10% whereas the QY of the films, which are likely less than the in-solution QYs due to aggregation effects, were not reported. Therefore, the research efforts need to be directed toward obtaining their efficient solids to use the plant extracts in solid-state lighting (SSL).

With this motivation, we studied the potential of water extract of *Peganum harmala*, also known as Syrian rust or African rue, for utilization in SSL. Phytochemically, this plant is rich in β-carbolines alkaloids such as harmine, harmaline, harmalol, and harman and quinazoline derivatives such as vasicine, vasicinone, and deoxyvasicinone[21,22] and has several applications, particularly for therapeutic purposes such as antibacterial, antifungal and antiviral effects [23–25], anti-tumour effect[26] and, insecticidal effect[12]. Although its blue-green fluorescence is known[27], the literature lacks sufficient information about its optical characterizations and their application in SSL. With this motivation, here, we prepared the water extract of *P. harmala* and characterized its optical features. Its quantum yield beyond 75% shows that this extract can be a good candidate for SSL if it can sustain these QYs at reasonable levels in solid form. For this purpose, we prepared solids of *Peganum harmala* by embedding the biomolecules into two crystalline matrices, sucrose and KCl, and into two fibrous matrices, cotton and paper. Then, we integrated the solid matrix having the highest QY with a UV LED chip to prepare color-converting LEDs.

## 2. Materials and Methods

### 2.1 Extraction of *P. harmala*

*P. harmala* was purchased from a local store in Kayseri, Türkiye. *P. harmala* seeds were ground using a mortar, and then 5 g of this powder was mixed with 50 mL of double distilled water (ddH$_2$O). The solution waited in a dark condition at room temperature for 24 h. For cleaning, the mixture was first filtered using Whatmann filter paper (No 1). Then, the solution was centrifuged at 5000 rpm for 10 min and filtered using a 0.2 µm cellulose acetate hydrophilic filter (Minisart, Sartorius). The aqueous solution of *P. harmala* seeds was stored at 4 °C for further use in our experiments.

## 2.2 Fabrication of fluorescent solid composites

### 2.2.1 Preparation of Crystals and Pellets of P. harmala

Prior to the experiments, fresh sucrose and salt stock solutions were prepared. For this purpose, 260 g and 170 g of sucrose and potassium chloride (KCl) were separately dissolved in 500 mL of ddH$_2$O. Solutions were filtered using a 0.2 µm hydrophilic filter to remove any impurities. For the crystallizations, *P. harmala* extract, stock sucrose or KCl solutions and ddH$_2$O were mixed at given volumes (**Table 1**) in a polystyrene petri dish without lid (60 mm). Samples were dried in a fume hood at room temperature.

**Table 1.** Solutions and their volumes used in the crystallization processes.

|  | Volume of *P. harmala* extract (mL) | Volume of stock sucrose solution (mL) | Volume of stock KCl solution (mL) | Volume of water (mL) |
|---|---|---|---|---|
| *P. harmala* in sucrose crystals | 1 mL | 2 mL | - | 2 mL |
| *P. harmala* in KCl crystals | 1 mL | - | 2 mL | 2 mL |
| Sucrose crystals (control group) | - | 2 mL | - | 3 mL |
| KCl crystals (control group) | - | - | 2 mL | 3 mL |

After the crystals were formed, samples were powdered using a mortar. Sucrose and KCl pellets (monoliths) were prepared by applying high pressure to these powders in a hydraulic press machine (Specac, Atlas Manual). Briefly, 125 mg of the powders were placed between two stainless steel disks, and a pressure of 0.75 GPa was applied using a hydraulic press for 10 min at room temperature.

*2.2.2 Preparation of fluorescent P. harmala cotton and fiber paper*

For embedding *P. harmala* extract into fibrous cellulose materials, cotton pads (Lure) and fiber papers (Munktell, 67 N grade) were employed. To prepare paper-based solids, firstly, small disk-shaped papers were cut at a diameter of ~1 cm from a general-purpose fiber drying paper. After that, 2 mL of *P. harmala* solution was transferred to a 60 mm polystyrene petri dish and disk-shaped papers were soaked into this solution and kept for 2 min. Afterwards, papers were taken out from the solution and kept in a fume hood until they were completely dry.

On the other hand, for cotton sample preparation, 2 mL of *P. harmala* solution was transferred to a 60 mm polystyrene petri dish. The cotton pad was completely soaked into this solution, and kept in a fume hood until they were completely dry.

As control groups, samples were prepared by following the same steps for cotton and paper hosts, except that only double distilled water (ddH$_2$O) instead of *P. harmala* solution was employed.

## 2.3 Optical and Structural Characterizations

Optical characterization of aqueous *P. harmala* extract was performed by absorption, photoluminescence, and quantum yield measurements. UV-Vis absorption spectrum of *P. harmala* extract was acquired with a UV-1800 UV-Vis spectrophotometer (Shimadzu) between 250-700 nm. The photoluminescence (PL) spectrum was recorded with an Agilent-Cary Eclipse fluorescence spectrophotometer at an excitation wavelength of 382 nm. The quantum yield (QY) of the solution was measured using a reference dye, Coumarin 153, which has a quantum yield of 54.2%. Briefly, the absorbance of the *P. harmala* extract and the Coumarin 153 dye was acquired. Then, the PL spectra of both extract and dye were measured at the excitation wavelength where both absorbance spectra cross. Using the PL data, the QY of the extract was calculated according to Equation 1:

$$\emptyset_{extract} = \frac{\int S_{extract}(\lambda)d\lambda}{\int S_{dye}(\lambda)d\lambda} \times \left(\frac{n_{extract}}{n_{dye}}\right)^2 \times \emptyset_{dye} \times 100\% \qquad \textbf{(Equation 1)}$$

where $\emptyset$ stands for the QY, $n$ corresponds to the refractive index, and S indicates the PL spectrum. Here, the extract is *P. harmala* extract, and the dye is Coumarin 153.

On the other hand, in solid-state, QYs of the sucrose pellet, KCl pellet, cotton, and fiber paper matrices were measured on a Hamamatsu absolute PL Quantum Yield Spectrometer C11347 using the integrating-sphere method ($\lambda_{exc}$ = 390 nm).

Structural characterizations were conducted using fluorescence microscopy and scanning electron microscopy (SEM) to understand the effect of morphological variations on the optical properties. Fluorescence microscopy images were recorded using a Nikon Eclipse Ni fluorescence microscope. Electron microscopy images were taken using ZEISS-GeminiSEM 300 Scanning Electron Microscopy at 3 kV or 5 kV. Prior to the imaging, 5 nm Au layer was sputtered on the sample.

## 2.4 Light-Emitting Diode (LED) Application

The paper sample having the highest QY was used for LED measurement. The paper was directly placed on a commercially available low-power LED having a peak emission wavelength of 400 nm. Prior to this process, a small volume of epoxy resin was added onto the LED to fix the sample on the LED. The luminance was recorded using a Hamamatsu integrating sphere and spectrometer. Colorimetric and photometric characterizations were carried out employing in-house written codes.

## 3. Results and Discussion

The green-fluorescent extract used in this study was obtained by brewing the powdered *P. harmala* seeds in double distilled water for 24 h (Figure 1a). To reveal the optical properties of the extract, the absorbance and fluorescence spectra of the aqueous *P. harmala* extract were examined (Figure 1b). Results demonstrated that *P. harmala* has a very prominent absorption peak at 365 nm. In the literature, although there are a couple of studies reporting the absorption spectra of green synthesized nanoparticles using *P. harmala* extracts as reducing agent[28,29], there is only one report about the absorption spectrum of *P. harmala* extract in water[30]. Interestingly, the absorbance spectrum reported by Azizi et al.[30] does not have any strong peaks but has weaker shoulders around 460 and 640 nm. On the other hand, our extract possesses a clear and strong absorbance peak of around 365 nm, which is similar to the report of Bhattacharjee *et al.* on harmaline[31]. The difference in the spectra between our extract and Azizi et al.[30] may stem from different extraction conditions employed. Furthermore, the similarity between our absorbance spectra and that of Bhattacharjee et al.[31] indicates that harmaline is the most prominent optically-active chemical in the extract.

To characterize the emission profile of our extract, we performed a photoluminescence (PL) measurement, and the results showed that our extract has a fluorescence peak at 484 nm when it is excited at 382 nm. Here, similar to the absorbance spectra, there is a strong similarity between the PL of our extract and the harmaline PL spectrum reported by Bhattacharjee et al.[31], which further supports our previous conclusion on the dominance of harmaline throughout the extract.

Quantum yield is one of the main parameters in SSL applications. Thus, we measured the QY of the *P. harmala* extract and found it as 75.6%. Such a high value makes it comparable to the levels that inorganic, toxic quantum dots or some fluorescent polymers provide[32,33]. In the literature, Hidalgo et al.[34] reported the QYs of norharman, harmane, harmalol, and harmol, which are the chemical compounds in *P. harmala* extract, as 16%, 17%, 39%, and 49%, respectively. At this point, it is worth highlighting that the QY of the extracts may be affected by the extraction procedure (e.g. solvent type, method, time, etc.) and purification processes, potentially leading to lower QYs in the study of Hidalgo *et. al.* than the one we measured.

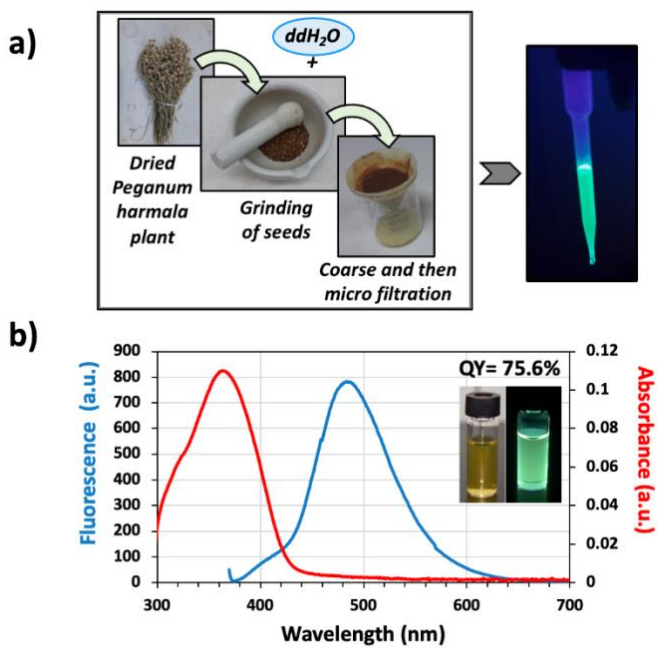

**Figure 1.** a) Representative image of extraction procedure from *P. harmala* seed. b) Absorbance and fluorescence data of *P. harmala* extract in water. QY data and the photos of *P. harmala* extract in water under daylight and UV light are also shown on the top right side.

A decrease in quantum yield during the transition from liquid to solid state is a significant problem in avoiding the use of fluorescent materials in SSL. To overcome this problem, our group previously demonstrated that when inorganic and organic nanoparticles are incorporated into crystalline matrices, the QYs and the emission stability of these fluorophores improve[32,35–37], making these nanomaterials attractive for SSL applications. However, the distribution of plant-based fluorophores and their optical properties in these matrices have remained unclear.

In light of these previous findings, we first approached the *P. harmala* extract crystallization in saturated sucrose and potassium chloride (KCl) solutions. Obtained crystal structures and their preparation details are provided in Figure 2 and Table 1, respectively. As seen in Figure 2 and magnified images of the crystals provided in Figure S1, KCl crystals tend to form small crystal grains separately and *P. harmala* extract is localized in some parts of these grains. However, as opposed to KCl crystal, sucrose formed a large and continuous one piece of crystal and covered the petri dish surface relatively homogeneously. Furthermore, the fluorescent molecules of the *P. harmala* extract are distributed more homogeneously within the whole sucrose crystal structure instead of localizing in some parts. The strong ionic forces in salt solution might be caused to the aggregation of luminescent molecules and prevent a homogenous distribution of the extract within the salt matrix. Additionally, although we used the same total volume (5 mL, Table 1) in both the control and *P. harmala* groups, we observed that *P. harmala* including groups covered a larger area on the dish surface than the control groups (Figure 2). The lowered cohesive forces and, thus, the lowered surface tension of water in the presence of extract have contributed to this observation.

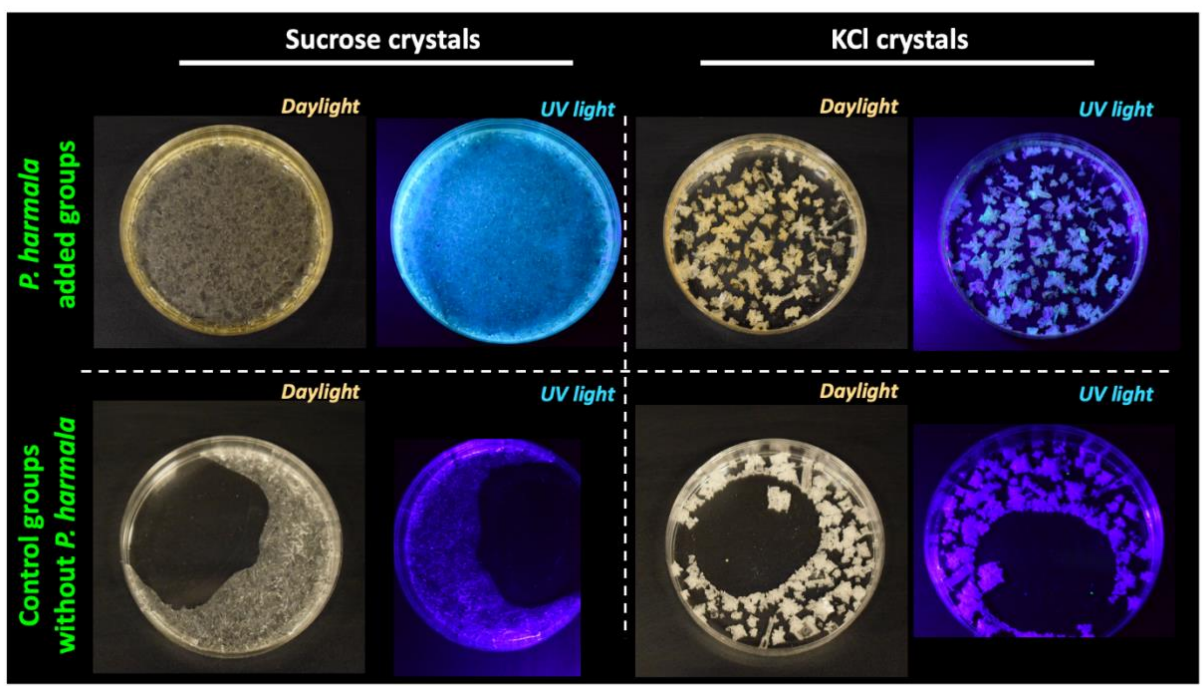

**Figure 2.** Images of sucrose and KCl crystals with and without *P. harmala* under ambient lighting and UV illumination.

Although *P. harmala* extract is transferred to the solid state in these crystals, the fabrication of a pellet (monoliths), which includes homogenously distributed fluorescent biomolecules and allows for easy handling, is crucial for LED studies. For this purpose, first, we ground these sucrose- and KCl-based macrocrystals into powder forms using a mortar. Then, we obtained the pellets having a diameter of ~1 cm by applying high pressure to these powders. The images of the powders and monoliths under daylight and UV light are presented in Figure 3. Consistent with Figure 2, green fluorophores aggregated at specific locations of the KCl pellet and represented a nonhomogeneous distribution. However, the homogenous distribution of the fluorophores which causes homogenous greenish emission was observed within the sucrose pellets (Figure 3, left-top).

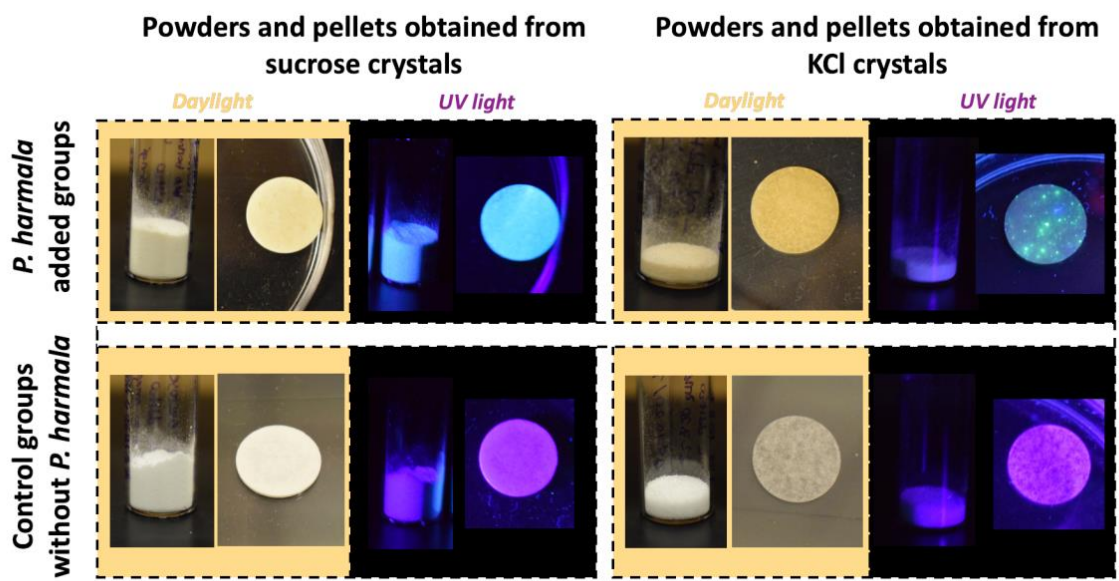

**Figure 3.** Images of powders obtained from ground crystals and solid pellet structures obtained from powders under daylight and UV light. For sucrose and KCl, control groups and *P. harmala* incorporated groups are presented together.

In addition to the crystalline matrices, we also investigated two different fibrous matrices. Using these materials, first, we aimed to avoid the aggregation of the fluorescent molecules, which is more likely to occur during drying and crystal formation. Secondly, the random distribution of *P. harmala* extract in different fiber structures and the effect of this distribution on the optical features were studied. In accordance with this, here we employed cellulose, which is one of the most abundant sustainable, renewable, and biocompatible natural polymers in the biosphere. Besides, thanks to their chemical structure, cellulose-based materials are known as exceptional adsorbants[38]. Considering their outstanding features, we immersed the fluorescent *P. harmala* extract into cellulose-based materials and compared them with our crystal-based host matrices. We chose cotton pad and fibrous drying paper as cellulose-based materials because of their easy accessibility and low cost. Briefly, cotton and fiber papers were soaked into *P. harmala* extract, and then they were dried (Figure 4). Although some orange shade was observed in some parts of the *P. harmala*-embedded cotton and paper in the daylight images, a homogenous emission was obtained under UV illumination.

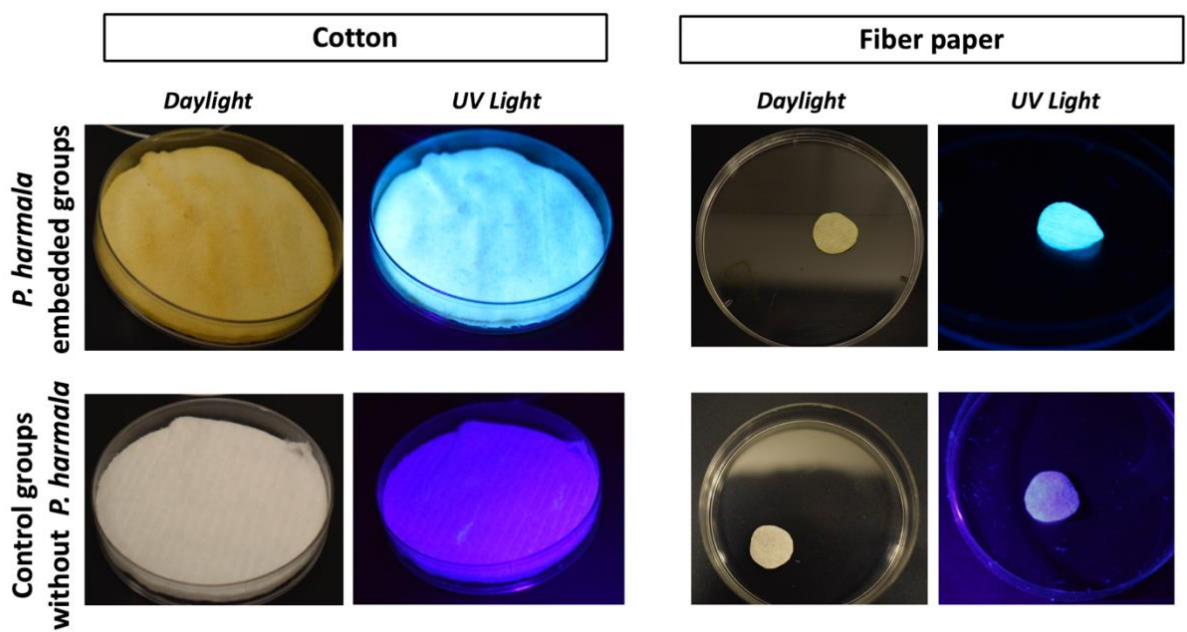

**Figure 4.** Images of cotton and fiber paper groups with and without *P. harmala* under daylight and UV light.

We also characterized the *P. harmala* embedded sucrose crystal, KCl crystal, cotton, and fibrous paper under a fluorescence microscope and scanning electron microscope (SEM) to understand the structure-extract distribution relation in detail (Figure 5). In the fluorescence microscope images, the sucrose pellet showed a cloudy structure and mostly homogeneous emission (Figure 5a). On the other hand, the appearance of the KCl pellet was not cloudy and the details were distinguishable. However, the green-emitting areas were nonhomogeneous across the monolith in contrast to the sucrose pellet (Figure 5b). When we looked closer at both sucrose and KCl monoliths, small spot-like structures were observed. These structures were more prominent in the sucrose monoliths. Interestingly, the borders of these structures are fluorescent instead of their core, as indicated by the red arrows in Figures 5a and b. This reveals that *P. harmala* extract surrounds the sucrose and salt molecules. In addition, the distribution and interconnection of these bead-like structures are more homogenous in sucrose monolith than KCl monolith. As seen from SEM images given in Figure 5e and f, both sucrose and KCl monoliths have a smooth surface and linear lines (shown with white arrows) occurring because of the applied pressure between two steel cylinders in the hydraulic press to obtain the monolith structure. On the other hand, in the magnified images given in the insets, we observed small grape-like structures coming together, which supports our observations in Figure 5a and b.

Compared to the crystal-based monoliths, cellulose-based cotton and fiber paper revealed fibrillar structures in both fluorescence and SEM images (Figure 5c,d,g,h). In the fluorescence images, "fiber paper" fibers appeared more flattened compared to cotton fibers as shown with red arrows in Figure 5d. To understand the structural details, we performed SEM imaging. As represented in Figures 5g and 5h, we observe a thicker and more flattened fiber structure in the paper in contrast to the cotton, which supports the images obtained by fluorescence microscope. Also, SEM images indicated that cotton fibers have a diameter of $16.8\pm4.6$ μm, and they appeared to have a more helical shape. On the other hand, in fiber paper, it seems like thinner fibers come together and form these thick fibers having a diameter of $28.9\pm8.6$ μm. Also, many short extensions between the fibers are visible in SEM images.

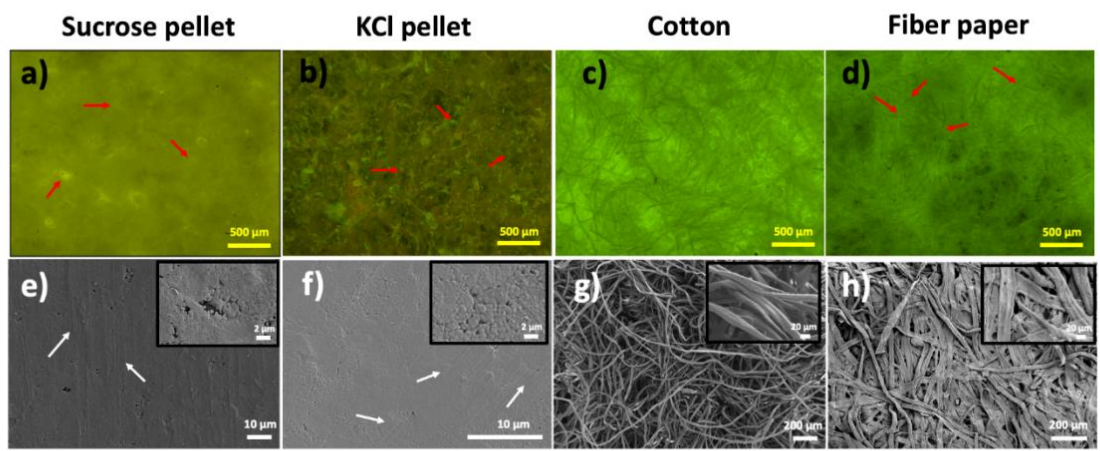

**Figure 5.** a) Fluorescence microscopy and b) SEM images of sucrose pellet, KCl pellet, cotton, and fiber paper samples, including *P. harmala* extract. (Scale bars of fluorescence images are 500 µm. Scale bars of SEM images are e-f: 10 µm and g-h: 200 µm. Scale bars of insets are e-f: 2 µm and g-h: 20 µm). Fluorescence microscope images were taken under an excitation wavelength of 450 nm.

Subsequent to structural characterization, the fluorescence quantum yield (QY) of all *P. harmala* samples was measured and presented in Figure 6. Our results revealed that the extract-embedded paper has the highest QY to be 44.7% decreased from 75.6% in the liquid state, which makes it suitable for a proof-of-concept LED application. Interestingly, although cotton also has a fibrous structure, the QY of this sample remained at 6.1%. There could be two explanations for this decreased quantum efficiency. The first one is that the absorption capacity of cotton is higher than paper, so it can absorb significantly more extracts. This means that more fluorescent biomolecules are found between solid cotton fibers and are close to each other leading to the quenching of the QY. The other possible explanation for the decreased QY is the difference of fiber structures between cotton and fiber paper. In the paper, wide, flat, and interconnected fibers increasing the surface area (Figure 5h) might also increase the distribution of the biomolecules after embedding, which may have caused less aggregation than the cotton. However, cotton possesses helical and thinner fibers (Figure 5g), possibly leading to a more serious aggregation and thus quenching.

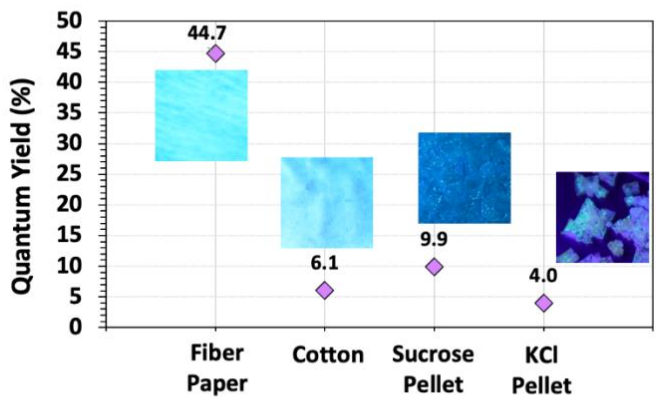

**Figure 6.** Quantum yields of *P. harmala* incorporated hosts.

Finally, we prepared a proof-of-concept LED with the fabricated sustainable bio-based color converters . Since we obtained the highest QY with the sample in paper host, we employed it as a biocompatible and easy-to-handle color converter on the LED chip emitting at ~400 nm. The representative image showing the fabrication of the LED set-up was given in Figure 7a, and the results are demonstrated in Figure 7b and c. The LED that we designed exhibited a blueish emission as quantified by the CIE 1931 color coordinates of (0.139,0.070). Although the extract itself has a cyan-green emission, the LED chip exciting the color-converting material shifted the color towards a more blueish region. We measured the luminous efficiency of our device as 21.9 lm/W$_{elect}$, which is relatively close to the performance of commercial epitaxially grown LED chips[39], showing the potential of plant extracts as color converters for SSL.

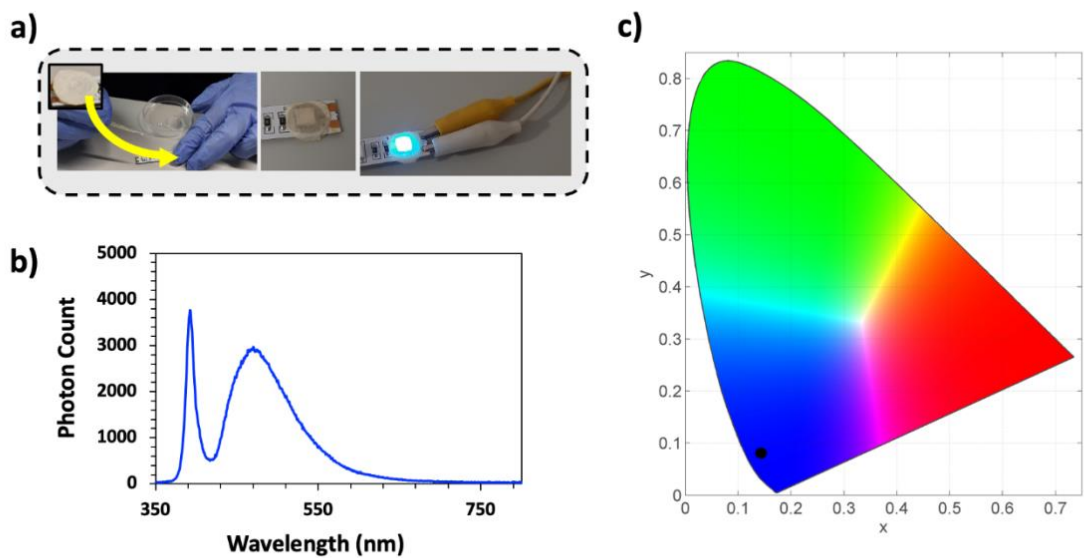

**Figure 7.** a) Representative image of the integration of the color-converting *P. harmala* embedded paper with the LED, b) LED emission spectrum under 5 mA current flow, and c) CIE 1931 color coordinates of the obtained LED on the chromaticity diagram.

**Conclusions**

In this study, we report a natural, cost-effective, easy-to-handle paper-based color converter obtained from fluorescent *P. harmala* extract, which can be a promising alternative to its counterparts in solid-state lighting. Here, to obtain color converters with relatively high QY in the solid state, we wrapped fluorescent biomolecules from plant extract into crystal-based and cellulose fiber-based matrices. As a crystal-based platform, we incorporated *P. harmala* extract into sucrose and KCl crystals. In parallel to these samples, we also studied cotton pads and drying paper (fiber paper) as cellulose fiber-based hosts to investigate their potential as a solid scaffold. Structural characterizations demonstrated that except KCl crystals, the extract is relatively homogenously distributed within the host matrices. However, QY measurements revealed that the highest quantum efficiency is observed in fiber paper at 44.7%, thanks to its structural advantages. In contrast, QYs of the other matrices become less than 10%. As a proof-of-concept demonstration, we presented the use of *P. harmala* extract embedded fiber paper as color converter on an LED possessing blue emission and exhibiting a luminous efficiency of 21.9 lm/$W_{elect}$ showing that *Peganum harmala* solids are excellent candidates as color converters. We believe that this study can pave the way for innovative new alternatives to the currently used color converters.

## AUTHOR INFORMATION

### Corresponding Authors

Zeliha Soran Erdem: zeliha.soranerdem@agu.edu.tr, Talha Erdem: erdem.talha@agu.edu.tr

### Author Contributions

The manuscript was written through contributions of all authors. All authors have given approval to the final version of the manuscript.

## ACKNOWLEDGEMENTS

ZSE and TE thank Abdullah Gül University for FFS funding.